\documentclass[onecollarge,natbib]{svjour2}
\bibpunct{[}{]}{;}{n}{}{,} 
\smartqed  
\usepackage{graphicx}
\journalname{Few Body Systems}

\usepackage{amssymb,amsmath,epsfig}

\renewcommand{\d}{\mathrm{d}}
\newcommand{\bm}[1]{{\bf #1}}

\newcommand{\mc}[1]{\mathcal{#1}}
\renewcommand{\sl}{\slashed}

\renewcommand{\d}{\mathrm{d}}
\newcommand{\sT}{{\scriptscriptstyle T}}
\newcommand{\pT}{\bm{p}_\sT}
\newcommand{\kT}{\bm{k}_\sT}
\newcommand{\qT}{\bm{q}_\sT}

\renewcommand{\Re}{\text{Re}}
\renewcommand{\Im}{\text{Im}}
\newcommand{\llll}{{\lambda_1 \lambda_2 \lambda_3 \lambda_4}}
\newcommand{\M}{M}

\begin{document}

\title{Using linear gluon polarization inside an unpolarized proton 
to determine the Higgs spin and parity
}

\author{Wilco J. den Dunnen}

\institute{Wilco J. den Dunnen \at
              Institute for Theoretical Physics, 
              Universit\"{a}t T\"{u}bingen, 
              Auf der Morgenstelle 14, D-72076 T\"{u}bingen, Germany\\
              \email{wilco.den-dunnen@uni-tuebingen.de}
}

\date{Received: date / Accepted: date}

\maketitle

\begin{abstract}

Gluons inside an unpolarized proton are in general linearly polarized
in the direction of their transverse momentum, rendering the LHC effectively
a polarized gluon collider.
This polarization can be utilized in the determination
of the spin and parity of the newly found Higgs-like boson.
We focus here on the determination of the spin using the azimuthal Collins-Soper
angle $\phi$ distribution.

\end{abstract}

\section*{}
In July 2012 the ATLAS and CMS collaborations announced the discovery
of a new resonance \cite{:2012gk,:2012gu} in their search for the Standard Model (SM) Higgs boson.
The current experimental challenge is the verification of all its
properties as predicted by the SM, in particular its spin and parity 
\cite{Aad:2013xqa,Chatrchyan:2012jja,ATLAS:2013mla,CMS:yva}.
The SM spin and parity prediction needs to be verified in all its decay channels independently.

In the diphoton decay channel the analysis method is based on measuring the
polar Collins-Soper (CS) angle $\theta$, 
which does not contain any information on the parity of the coupling, 
nor can it be used to distinguish between all possible spin-2 coupling scenarios
\cite{Choi:2002jk,Gao:2010qx,Bolognesi:2012mm,Choi:2012yg,Ellis:2012jv}.
We propose a measurement of the Higgs transverse momentum distribution as a way
to determine its parity and a measurement of the \emph{azimuthal} CS angle $\phi$ as 
an additional way to determine its spin and to distinguish between the different
spin-2 coupling possibilities \cite{Boer:2011kf,denDunnen:2012ym,Boer:2013fca}.

The underlying principle of these methods, relies on the fact that gluons are
linearly polarized in the direction of their transverse momentum when
extracted from an unpolarized proton.
This effect is, as far as we know, not taken into account in event generators,
which generate transverse momentum by parton showers that
leave the gluons unpolarized and consequently does not show up in `standard'
analyses.
Using the framework of Transverse Momentum Dependent (TMD) factorization 
one can systematically take into account partonic transverse momentum
and polarization.

\section*{Transverse Momentum Dependent factorization}

In the framework of Transverse Momentum Dependent factorization, 
the full $pp\to \gamma\gamma X$ cross section is split into a partonic
$gg\to \gamma\gamma$ cross section and two TMD gluon correlators, that describe the 
distribution of gluons inside a proton as a function of not only its momentum along the
direction of the proton, but also transverse to it.
More specifically, the differential cross section for the inclusive 
production of a photon pair from gluon-gluon fusion is written as 
\cite{Ji:2005nu,Sun:2011iw,Ma:2012hh},
\begin{equation}\label{eq:factformula}
\frac{\d\sigma}{\d^4 q \d \Omega}
  \propto 
  \int\!\! \d^{2}\pT \d^{2}\kT
  \delta^{2}(\pT + \kT - \qT)
  \mc{M}_{\mu\rho \kappa\lambda}
  \left(\mc{M}_{\nu\sigma}^{\quad\kappa\lambda}\right)^*
  \\
  \Phi_g^{\mu\nu}(x_1,\pT,\zeta_1,\mu)\,
  \Phi_g^{\rho\sigma}(x_2,\kT,\zeta_2,\mu),
\end{equation}
with the longitudinal momentum fractions $x_1={q\cdot P_2}/{P_1\cdot P_2}$
and $x_2={q\cdot P_1}/{P_1\cdot P_2}$, $q$ the momentum of the photon pair, 
$\mc{M}$ the $gg\to \gamma\gamma$ partonic hard scattering matrix element
and $\Phi$ the following gluon TMD correlator in an unpolarized proton,
\begin{align}\label{eq:TMDcorrelator}
\Phi_g^{\mu\nu}(x,\pT,\zeta,\mu) 
&\equiv
	      \int \frac{\d(\xi\cdot P)\, \d^2 \xi_\sT}{(x P\cdot n)^2 (2\pi)^3}
	      e^{i ( xP + p_\sT) \cdot \xi}
	      \left\langle P \left| F_a^{n\nu}(0)
	      \left(\mc{U}_{[0,\xi]}^{n[\text{--}]}\right)_{ab} F_b^{n\mu}(\xi)
	      \right|P \right\rangle \Big|_{\xi \cdot P^\prime = 0}\nonumber\\
&=	-\frac{1}{2x} \bigg \{g_\sT^{\mu\nu} f_1^g(x,\pT^2,\zeta,\mu)
	-\bigg(\frac{p_\sT^\mu p_\sT^\nu}{M_p^2}\,
	{+}\,g_\sT^{\mu\nu}\frac{\pT^2}{2M_p^2}\bigg)
	h_1^{\perp\,g}(x,\pT^2,\zeta,\mu) \bigg \} + \text{HT},
\end{align}
with $p_{\sT}^2 = -\pT^2$ and $g^{\mu\nu}_{\sT} = g^{\mu\nu}
- P^{\mu}P^{\prime\nu}/P{\cdot}P^\prime - P^{\prime\mu}P^{\nu}/P{\cdot}P^\prime$,
where $P$ and $P^\prime$ are the momenta of the colliding protons and $M_p$ their mass.
The gauge link $\mc{U}_{[0,\xi]}^{n[\text{--}]}$ for this process arises 
from initial state interactions. 
It runs from $0$ to $\xi$ via minus infinity along the direction $n$, 
which is a time-like dimensionless four-vector with no transverse 
components such that $\zeta^2 = (2n{\cdot}P)^2/n^2$. 

In principle, Eqs.\ \eqref{eq:factformula} and \eqref{eq:TMDcorrelator} also contain
soft factors, but with the appropriate choice of $\zeta$ (of around 1.5 times $\sqrt{s}$), 
one can neglect their contribution, at least up to next-to-leading order 
\cite{Ji:2005nu,Ma:2012hh}.
To avoid the appearance of large logarithms in ${\cal M}$, 
the renormalization scale $\mu$ needs to be of order $M_h$.

The second line of Eq.\ \eqref{eq:TMDcorrelator} contains the parametrization
of the TMD correlator in the conventions of Ref.\ \cite{Mulders:2000sh},
where $f_1^g$ is the unpolarized gluon distribution and
$h_1^{\perp\,g}$ the linearly polarized gluon distribution.
The Higher Twist (HT) terms only give power suppressed contributions at small transverse momentum.

The effect of the linearly polarized gluon distribution is such that,
for positive values, the probability of finding a gluon with linear polarization
along its transverse momentum is larger than the probability of
finding it perpendicular to it.
For negative values, this is reversed. 
Full gluon polarization corresponds to $h_1^{\perp g}$ saturating its upper bound, i.e.,
$|h_1^{\perp g}| = 2M_p^2/\pT^2 f_1^g$ \cite{Mulders:2000sh}.

\section*{General structure of the cross section}

The general structure of the differential cross section for the process $pp\to \gamma\gamma X$
follows from Eq.\ \eqref{eq:factformula} and \eqref{eq:TMDcorrelator} and can be written as
(cf.\ Ref.\ \cite{Qiu:2011ai})
\begin{multline}\label{eq:genstruc}
\frac{\d\sigma}{\d Q \d Y \d^2 \qT\, \d\cos\theta\, \d\phi} 
  \propto 
	F_1\, \mc{C} \left[f_1^gf_1^g\right]
	+ F_2\,	\mc{C} \left[w_2\, h_1^{\perp g}h_1^{\perp g}\right]
	+ F_3\, 	\mc{C} \left[w_3 f_1^g h_1^{\perp g}
				  + (x_1 \leftrightarrow x_2) \right]\cos(2\phi)\\
	+ F_3^{\prime}\, \mc{C} \left[w_3 f_1^g h_1^{\perp g} 
				  - (x_1 \leftrightarrow x_2) \right]\sin(2\phi)
	+ F_4\,	\mc{C} \left[w_4\, h_1^{\perp g}h_1^{\perp g}\right]\cos(4\phi),
\end{multline}
up to corrections that are $\qT^2/Q^2$ suppressed at small $\qT$.
The cross section is differential in $Q$, $Y$ and $\qT$, which are the invariant mass,
rapidity and transverse momentum of the pair in the lab frame
and in the Collins-Soper angles $\theta$ and $\phi$.
The latter two are defined as the polar and azimuthal angle
in the Collins-Soper frame \cite{Collins:1977iv},
which is the diphoton rest frame with the $\hat{x}\hat{z}$-plane spanned by the 3-momenta 
of the colliding protons and the $\hat{x}$-axis set by their bisector. 
The convolution $\mc{C}$ is defined as
\begin{equation}
\mathcal{C}[w\, f\, g] \equiv \int\! \d^{2}\pT\int\! \d^{2}\kT\,
  \delta^{2}(\pT+\kT-\bm q_{\sT})
  w(\pT,\kT)\, f(x_{1},\pT^{2})\, g(x_{2},\kT^{2}),
\end{equation}
in which the longitudinal momentum fractions are given in 
the aforementioned kinematical variables by
\begin{equation}
x_{1,2} = e^{\pm Y} \sqrt{\frac{Q^2 + \qT^2}{s}}. 
\end{equation}
The weights that appear in the convolutions are defined as
\begin{align}
 w_2		&\equiv \frac{2 (\kT{\cdot}\pT)^2 - \kT^2 \pT^2 }{4 M_p^4},\nonumber\\
 w_3 		&\equiv \frac{\qT^2\kT^2 - 2 (\qT{\cdot}\kT)^2}{2 M_p^2 \qT^2},\nonumber\\
 w_4		&\equiv  2\left[\frac{\pT{\cdot}\kT}{2M_p^2} - 
		\frac{(\pT{\cdot}\qT) (\kT{\cdot}\qT)}{M_p^2\qT^2}\right]^2 -\frac{\pT^2\kT^2 }{4 M_p^4},
\end{align}
and the $F_i$ factors can be expressed as,
\begin{align}\label{eq:Fdefs}
F_1
			&= \sum_{\lambda_1,\lambda_2,\lambda_3,\lambda_4} 
			\M^{\llll} \left(\M^{\llll} \right)^*, \nonumber\\
F_2
			&= 2\sum_{\lambda_3,\lambda_4} \Re\left[\M^{++\lambda_3\lambda_4}
			\left(\M^{--\lambda_3\lambda_4}\right)^*\right], \nonumber\\
F_3
			&= \sum_{\lambda,\lambda_3,\lambda_4} 
			\Re\left[
			\M^{\lambda-\lambda_3\lambda_4} 
			\left(\M^{\lambda+\lambda_3\lambda_4}\right)^*
			+ \M^{+\lambda\lambda_3\lambda_4} 
			\left(\M^{-\lambda\lambda_3\lambda_4}\right)^*
			\right], \nonumber\\
F_3^{\prime}
			&= \sum_{\lambda,\lambda_3,\lambda_4} 
			\Im\left[
			\M^{\lambda-\lambda_3\lambda_4} 
			\left(\M^{\lambda+\lambda_3\lambda_4}\right)^*
			- \M^{+\lambda\lambda_3\lambda_4} 
			\left(\M^{-\lambda\lambda_3\lambda_4}\right)^*
			\right], \nonumber\\
F_4
			&= 2\sum_{\lambda_3,\lambda_4} \Re\left[\M^{+-\lambda_3\lambda_4}
			\left(\M^{-+\lambda_3\lambda_4}\right)^*\right],
\end{align}
in terms of the $gg\to\gamma\gamma$ helicity amplitudes,
that are defined by
\begin{equation}\label{eq:helampdef}
 M^\llll \equiv \epsilon_\mu^{\lambda_1}(p)^* \epsilon_\nu^{\lambda_2}(k)^*
	      M^{\mu\nu\rho\sigma} \epsilon_\rho^{\lambda_3}(q_1) \epsilon_\sigma^{\lambda_4}(q_2),
\end{equation}
in terms of the covariant polarization vectors,
\begin{align}
\epsilon_{\lambda}^{\mu}(p)	= \epsilon_{-\lambda}^{\mu}(k)
				&= \frac{1}{2 \sqrt{\Delta}}
				\left(\epsilon^{\mu p k q_1} - i\lambda L^\mu\right),\nonumber\\
\epsilon_{\lambda}^{\mu}(q_1) 	= \epsilon_{-\lambda}^{\mu}(q_2)
				&=\frac{1}{2 \sqrt{\Delta}}
				\left(\epsilon^{\mu p k q_1} - i\lambda K^\mu\right),
\end{align}
in which
\begin{align}
\Delta &\equiv (k\cdot q_1)(p\cdot q_1) (p\cdot k) \nonumber\\
L^\mu &\equiv (k\cdot q_1) p^\mu + (p\cdot q_1) k^\mu - (p\cdot k) q_1^\mu,\nonumber\\
K^\mu &\equiv (k\cdot q_1) p^\mu - (p\cdot q_1) k^\mu + (p\cdot q_1 - k\cdot q_1)q_1^\mu.
\end{align}

\section*{Partonic cross section}
We will consider the partonic process $gg\to X_{i}\to \gamma\gamma$,
where $X_{i}$ can either be a spin-0, spin-1 or spin-2 boson.
The helicity amplitudes, as defined in Eq.\ \eqref{eq:helampdef},
will be given in the following matrix notation,
\begin{equation}
 M_{X_i} = 
  \left(
\begin{array}{ccccc}
 M^{+-+-}
 & M^{+-++}
 & M^{+---}
 & M^{+--+} \\
 M^{+++-}
 & M^{++++}  
 & M^{++--}
 & M^{++-+} \\
 M^{--+-}
 & M^{--++} 
 & M^{----} 
 & M^{---+} \\
 M^{-++-}
 & M^{-+++}
 & M^{-+--}
 & M^{-+-+} \\
\end{array}
\right).
\end{equation}

\section*{Spin-0}
For a spin-0 boson the helicity amplitudes read
\begin{equation}
 M_{X_0} = 
  \left(
\begin{array}{ccccc}
 0
 & 0
 & 0
 & 0 \\
 0
 &\,\, V_{gg0}^{++} {V_{\gamma\gamma0}^{++}}^*\,\,
 &\,\, V_{gg0}^{++} {V_{\gamma\gamma0}^{--}}^*\,\,
 & 0 \\
 0
 & V_{gg0}^{--} {V_{\gamma\gamma0}^{++}}^* 
 & V_{gg0}^{--} {V_{\gamma\gamma0}^{--}}^* 
 & 0 \\
 0
 & 0
 & 0
 & 0 \\
\end{array}
\right),
\end{equation}
in which $V_{gg0}^{\pm\pm}$ and $V_{\gamma\gamma0}^{\pm\pm}$ are the 
$g g X_0$ and $\gamma\gamma X_0$ helicity vertices.
We will assume equal coupling to gluons and photons and
express the helicity vertices in the conventions of Refs.\ \cite{Gao:2010qx} 
and \cite{Bolognesi:2012mm}, i.e.,
\begin{align}
 V_{gg 0}^{++} &= V_{\gamma\gamma 0}^{++} = a_1 + \frac{1}{2}i a_3, \nonumber\\
 V_{gg 0}^{--} &= V_{\gamma\gamma 0}^{--} = a_1 - \frac{1}{2}i a_3,
\end{align}
up to a constant factor that will be irrelevant for us as we will only be interested
in distributions and not the absolute size of the cross section.
In this parametrization of the vertex, 
the following non-zero $F$ factors in Eq.\ \eqref{eq:genstruc} are obtained
\begin{align}\label{eq:Fspin0}
 F_1 &= (4|a_1|^2 + |a_3|^2)^2, \nonumber\\
 F_2 &= (4|a_1|^2 + |a_3|^2)(4|a_1|^2 - |a_3|^2),
\end{align}
again up to a constant factor.

\section*{Spin-1}
For a spin-1 boson the helicity amplitudes would read
\begin{equation}
 M_{X_1} = 
  \left(
\begin{array}{ccccc}
 0
 & 0
 & 0
 & 0 \\
 0
 & - V_{gg1}^{++} {V_{\gamma\gamma1}^{++}}^* \cos\theta 
 & - V_{gg1}^{++} {V_{\gamma\gamma1}^{--}}^* \cos\theta 
 & 0 \\
 0
 &\,\, - V_{gg1}^{--} {V_{\gamma\gamma1}^{++}}^* \cos\theta\,\,
 &\,\, - V_{gg1}^{--} {V_{\gamma\gamma1}^{--}}^* \cos\theta\,\, 
 & 0 \\
 0
 & 0
 & 0
 & 0 \\
\end{array}
\right),
\end{equation}
which has the following behavior under interchange of either
initial or final state particles, $M_{X_1}(\theta) = -M_{X_1}(\theta + \pi)$.
However, for identical particles in either the initial our final state,
one should have $M_{X_1}(\theta) = M_{X_1} (\theta + \pi)$.
This implies that, for identical particles, $V_1^{++} = V_1^{--} = 0$ and
that this partonic channel is thus forbidden, in accordance with the
Landau-Yang theorem \cite{Landau:1948kw,Yang:1950rg}.
In case of non-identical particles, the behavior of a spin-1
resonance would be equal to that of a spin-0 boson, 
but with a characteristic $\cos^2\theta$ dependence of the cross section.

\section*{Spin-2}
For a spin-2 boson the helicity amplitudes read
\begin{multline}
 M_{X_2} = \\
  \left(
\begin{array}{ccccc}
 V_{gg2}^{+-} {V_{\gamma\gamma 2}^{+-}}^* \cos ^4\left(\frac{\theta }{2}\right) 
 &V_{gg2}^{+-} {V_{\gamma\gamma 2}^{++}}^*  \sqrt{\frac{3}{8}} \sin ^2(\theta ) 
 &V_{gg2}^{+-} {V_{\gamma\gamma 2}^{--}}^* \sqrt{\frac{3}{8}} \sin ^2(\theta ) 
 &V_{gg2}^{+-} {V_{\gamma\gamma 2}^{-+}}^*  \sin ^4\left(\frac{\theta }{2}\right) \\
 V_{gg2}^{++} {V_{\gamma\gamma 2}^{+-}}^* \sqrt{\frac{3}{8}} \sin ^2(\theta ) 
 & V_{gg2}^{++} {V_{\gamma\gamma 2}^{++}}^* \frac{3 \cos (2 \theta )+1}{4} 
 & V_{gg2}^{++} {V_{\gamma\gamma 2}^{--}}^* \frac{3 \cos (2 \theta )+1}{4} 
 & V_{gg2}^{++} {V_{\gamma\gamma 2}^{-+}}^* \sqrt{\frac{3}{8}} \sin ^2(\theta ) \\
 V_{gg2}^{--} {V_{\gamma\gamma 2}^{+-}}^* \sqrt{\frac{3}{8}} \sin ^2(\theta ) 
 & V_{gg2}^{--} {V_{\gamma\gamma 2}^{++}}^* \frac{3 \cos (2 \theta )+1}{4} 
 & V_{gg2}^{--} {V_{\gamma\gamma 2}^{--}}^* \frac{3 \cos (2 \theta )+1}{4} 
 & V_{gg2}^{--} {V_{\gamma\gamma 2}^{-+}}^* \sqrt{\frac{3}{8}} \sin ^2(\theta ) \\
 V_{gg2}^{-+} {V_{\gamma\gamma 2}^{+-}}^* \sin ^4\left(\frac{\theta }{2}\right) 
 & V_{gg2}^{-+} {V_{\gamma\gamma 2}^{++}}^*  \sqrt{\frac{3}{8}} \sin ^2(\theta ) 
 & V_{gg2}^{-+} {V_{\gamma\gamma 2}^{--}}^* \sqrt{\frac{3}{8}} \sin ^2(\theta ) 
 & V_{gg2}^{-+} {V_{\gamma\gamma 2}^{-+}}^* \cos ^4\left(\frac{\theta }{2}\right) \\
\end{array}
\right),
\end{multline}
in which $V_{gg 2}^{\pm\pm}$ and $V_{\gamma\gamma 2}^{\pm\pm}$ are the 
$g g X_2$ and $\gamma\gamma X_2$ helicity vertices.
We will assume equal vertices for gluons and photons,
and express them in the conventions of Refs.\ \cite{Gao:2010qx} and \cite{Bolognesi:2012mm},
i.e.,
\begin{align}
 V_{gg 2}^{+-} &= V_{gg 2}^{-+} = V_{\gamma\gamma 2}^{+-} = V_{\gamma\gamma 2}^{-+} = c_1, \nonumber\\
 V_{gg 2}^{++} &= V_{\gamma\gamma 2}^{++} = \frac{1}{\sqrt{6}}[c_1 + 4 c_2 - i 2 (c_5 - 2 c_6)], \nonumber\\
 V_{gg 2}^{--} &= V_{\gamma\gamma 2}^{--} =  \frac{1}{\sqrt{6}}[c_1 + 4 c_2 + i 2 (c_5 - 2 c_6)],
\end{align}
up to a constant factor.
In the remainder, $c_6$ will be dropped as it can, in the coupling to massless particles,
not be distinguished from $c_5$.
The following non-zero $F$ factors in Eq.\ \eqref{eq:genstruc} are obtained
\begin{align}\label{eq:Fspin2}
 F_1 &= 18 A^+ |c_1|^2 \sin^4\theta +  {A^+}^2\! \left( 1 - 6\cos^2\theta + 9\cos^4\theta \right)
	+ 9 |c_1|^4 \left(1 + 6 \cos^2\theta + \cos^4\theta \right),\nonumber\\
 F_2 &= 9\, A^- |c_1|^2 \sin^4\theta +  A^-A^+\left( 1 - 6\cos^2\theta + 9\cos^4\theta \right),\nonumber\\
 F_3 &= 6\, B^-\, \left[ A^+( 3 \cos^2\theta - 1) + 3 |c_1|^2 (\cos^2\theta + 1) \right] \sin^2\theta,\nonumber\\
 F_3^\prime &= 12\, \Re(c_1 c_5^*)\, 
		\left[ A^+( 3 \cos^2\theta - 1) + 3 |c_1|^2 (\cos^2\theta + 1) \right] \sin^2\theta,\nonumber\\
 F_4 &= 18\, |c_1|^2\, \left[B^+ + 2 |c_5|^2\right] \sin^4\theta,
\end{align}
up to a constant factor and where we have defined 
$A^\pm\equiv |c_1+4c_2|^2\pm 4|c_5|^2$, $B^\pm\equiv |c_1 + 2 c_2|^2\pm 4|c_2|^2$.

\section*{Results}

We will concentrate here on the CS $\phi$ distribution and refer for 
predictions on the transverse momentum distribution to our Ref.\ \cite{Boer:2013fca}.
In addition to our Ref.\ \cite{Boer:2013fca} we will also consider
forward Higgs production, which can be used to search for a $CP$-violating
spin-2 Higgs coupling. 

To make numerical predictions we will use the the various coupling scenarios
that are defined in Ref.\ \cite{Bolognesi:2012mm}, 
to which we will add $2_{h^\prime}^+$, $2_{h^{\prime\prime}}^+$
and $2_{\text{CPV}}$.
The first two scenarios that we add serve as an example of two spin-2 
coupling hypotheses that are indistinguishable in the $\theta$ distribution, 
but \emph{do} have a different $\phi$ distribution.
The last one serves as an example of a spin-2 coupling
that violates $CP$ symmetry.
The scenarios are summarized in Table \ref{tab:bmscenarios}.

\begin{table}[htb]
 \begin{tabular}{cccrrrrrr}
  	&$0^+$ &$0^-$ &$2_m^+$ &$2_h^+$ &$2_{h^\prime}^+$ &$2_{h^{\prime\prime}}^+$ &$2_{h}^-$ &$2_{\text{CPV}}$\\
 \hline
 \hline
 $a_1$		&1 &0 &- &- &- &- &- &-\\
 $a_3$		&0 &1 &- &- &- &- &- &-\\
 $c_1$		&- &- &1 &0 &1 &1 &0 &1\\
 $c_2$		&- &- &$-\frac{1}{4}$ &1 &1 &$-\frac{3}{2}$ &0 &0 \\
 $c_5$ 		&- &- &0 &0 &0 &0 &1 &5\\
 \end{tabular} 
\caption{Various spin, parity and coupling scenarios.}\label{tab:bmscenarios}
\end{table}

To make numerical predictions we use the same Ansatz for 
$f_1^g$ as described in our Ref.\ \cite{Boer:2013fca}.
For forward Higgs production, we now also need $f_1^g$ at
$x=m_H/\sqrt{s}\exp[\pm Y]$, whereas in the aforementioned
Ref.\ $f_1^g$ was only considered at $x=m_H/\sqrt{s}$.
Within the range of rapidity that we will consider, $|Y|\leq 1$,
we will approximate the shape of $f_1^g$
to be independent of $x$, i.e., in the language of Ref.\ \cite{Boer:2013fca}
only $A_0$ depends on $x$.

The linearly polarized gluon distribution will be expressed in terms of
the degree of polarization $\mc{P}$ and the unpolarized gluon distribution, i.e.,
$h_1^{\perp g} = \mc{P}\, 2M_p^2/\pT^2 f_1^g$.
The degree of polarization $\mc{P}$, as function of $x$ and $\pT$, 
will be calculated using the methods described in Ref. \cite{Dunnen:2013zua}.

\begin{figure}[h!]
\centering
\includegraphics[width=7.4cm]{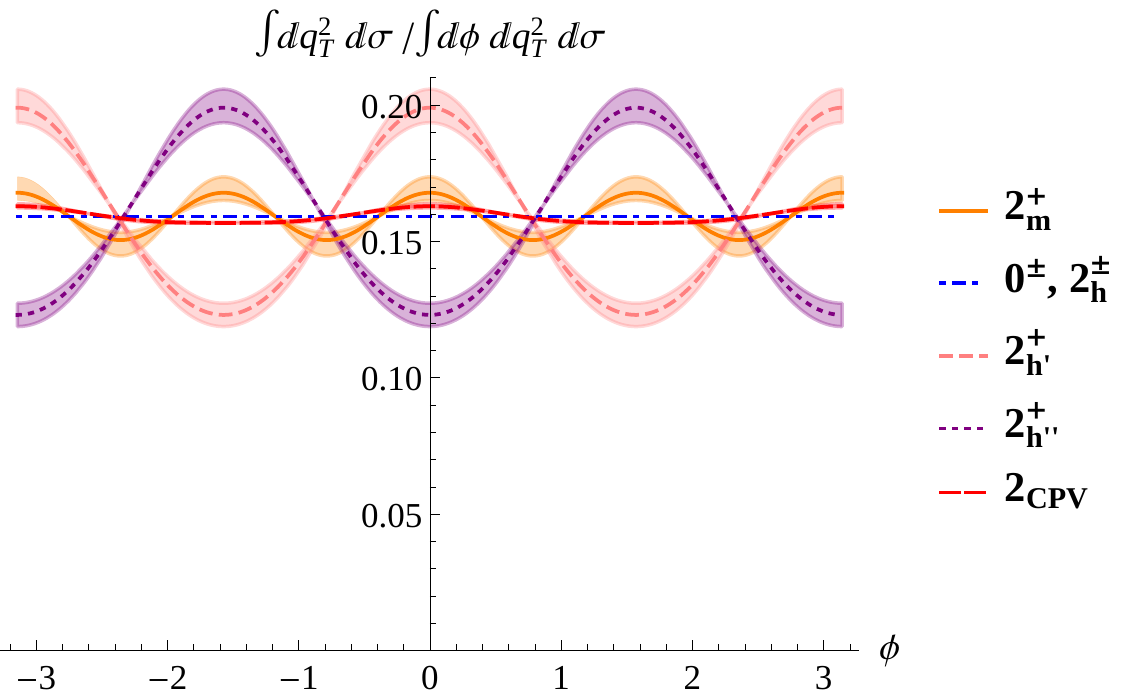}
\hspace{0.5cm}
\includegraphics[width=7.4cm]{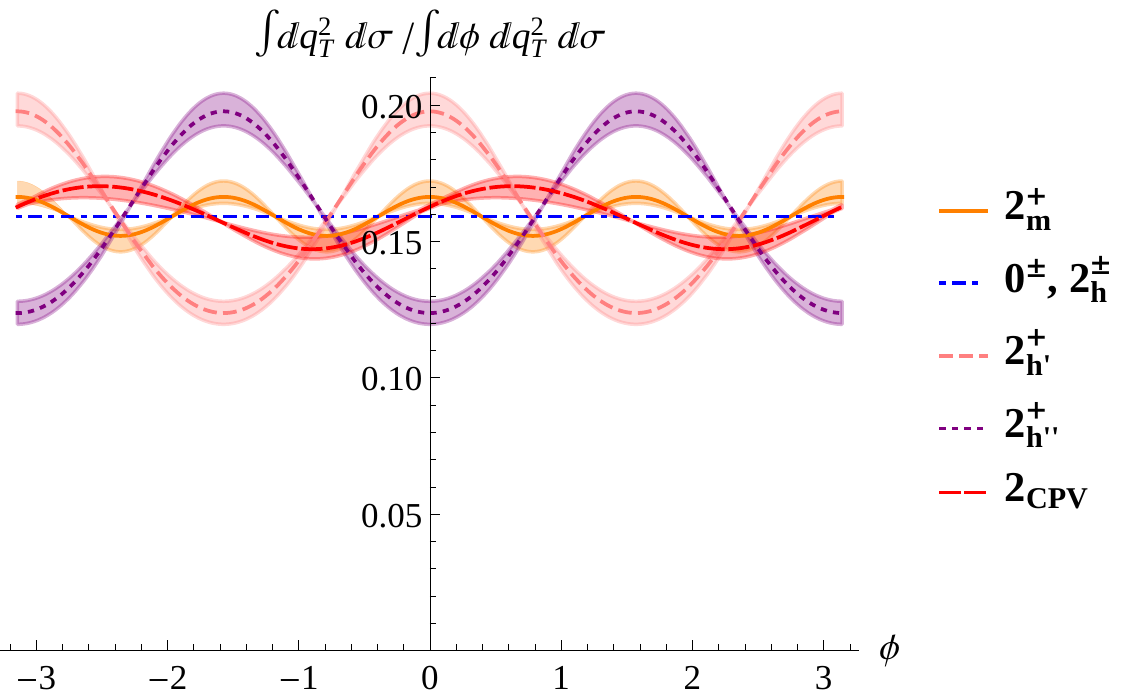}
\caption{Plot of the CS $\phi$ distribution
in the process $pp\to X_i X \to \gamma\gamma X$ for a 125 GeV resonance $X_i$
at a center of mass energy of 8 TeV, for various $X_i$ coupling scenarios
at $\theta=\pi/2$ and  at $Y=0$ (left) and $Y=1$ (right).
The upper limit on the $q_\sT$ integration has been chosen as $M_h/2$.
The shaded area is due to the uncertainty in the degree of polarization.}
\label{fig:distrs}
\end{figure}


In Figure \ref{fig:distrs} we show our prediction for the CS $\phi$ distribution
for central and forward Higgs production. 
The backward distribution can be obtained from the forward one by replacing $\phi\to -\phi$.
From the plot we can see that various spin-2 coupling scenarios
produce non-isotropic $\phi$ distributions.
The difference from the isotropic spin-0 distribution is of such a size that,
with the given collected data set, it might well
be possible to put constraints on various spin-2 coupling hypotheses.
Especially the benchmark scenarios $2_{h^\prime}$ and $2_{h^{\prime\prime}}$
are significantly different from the spin-0 scenario and could therefore
relatively easily be excluded.
The $2_{\text{CPV}}$ benchmark scenario displays a characteristic
\emph{asymmetric} $\phi$ distribution in the forward region that 
can \emph{only} be caused by a $CP$-violating coupling.

\section*{Conclusion}
We have presented the CS angle $\phi$ distribution in the process 
$pp\to X_i X \to \gamma\gamma X$, for $X_i$ a spin-0 and spin-2
boson with generic couplings, taking into account the fact that gluons inside an unpolarized
proton are partially linearly polarized.
Numerical predictions of the $\phi$ distribution show that various spin-2 coupling 
scenarios differ substantially from the isotropic spin-0 prediction, 
to an extent that a measurement of this distribution, based on the current data set
collected by ATLAS and CMS, might already be enough to exclude these scenarios.

\begin{acknowledgements}
This work was supported in part by the German Bundesministerium f\"{u}r Bildung und Forschung (BMBF),
grant no. 05P12VTCTG.
\end{acknowledgements}

\end{document}